\documentclass{revtex4}
\usepackage{amssymb}
\usepackage{amsmath}

\input{tcilatex}
\begin{document}

\title{Quantization of \ Higher Dimensional Linear Dilaton Black Hole
Area/Entropy From Quasinormal Modes}
\author{I. Sakalli}
\email{izzet.sakalli@emu.edu.tr}
\affiliation{Department of Physics, Eastern Mediterranean University, G. Magusa, North
Cyprus, via Mersin 10, Turkey.}

\begin{abstract}
The quantum spectra of area and entropy of higher dimensional linear dilaton
black holes in various theories via the quasinormal modes method are
studied. It is shown that quasinormal modes of these black holes can reveal
themselves when a specific condition holds. Finally, we obtain that a higher
dimensional linear dilaton black hole has equidistant area and entropy
spectra, and both of them are independent on the spacetime dimension.
\end{abstract}

\maketitle

\section{Introduction}

Inspired by Giddings and Strominger \cite{Giddings}, Cl\'{e}ment et al. \cite%
{Clement} proposed a new four-dimensional static spherically symmetric
solution describing a black hole in a linear dilaton background so-called
linear dilaton back hole (LDBH). The LDBHs of \cite{Clement} are exact
solutions to Einstein-Maxwell-Dilaton (EMD) theory. In general, LDBHs
represent non-asymptotically flat (NAF) spacetimes and break all
supersymmetries. Since the LDBHs are NAF spacetimes, their physical mass can
be computed by the Brown-York formalism \cite{BrownYork}, and shown to
satisfy the first law of thermodynamics consistent with the definition of
the Hawking temperature and of the geometric entropy, which is the quarter
of the horizon area \cite{Hawking1,Hawking2}. The most remarkable features
of the LDBHs are that the temperature is constant (representing an
isothermal process) and the mass is independent of the temperature so that
the heat capacity becomes zero.

A long time ago, higher dimensional generalization of the LDBHs was
inapparently\ given in EMD theory by Chan et al. \cite{highLDBHemd} as some
limit of extremal dilaton black holes \cite{Maeda}. Recently, it has been
shown that higher dimensional static LDBH solutions exist in
Einstein-Yang-Mills-Dilaton (EYMD) \cite{highLDBHeymd} and
Einstein-Yang-Mills-Born-Infeld-Dilaton (EYMBID) \cite{highLDBHeymbid}
theories. A study of the thermodynamic properties of these black holes has
also been done \cite{MSH}. Surprisingly, however, there are no detailed
studies of the entropy/area spectra and the quasinormal modes (QNMs -- the
characteristic ringing frequencies) of the higher dimensional LDBHs. The
motivation of this paper is to fill such a gap.

The quantum properties of black holes have attracted much attention for many
years. Firstly, Bekenstein \cite{Bek1,Bek2,Bek3,Bek4,Bek5} presumed that the
black hole horizon area and entropy ought to be quantized. Bekenstein
proposed that the black hole horizon area is an adiabatic invariant, and has
a discrete and equally spaced spectrum

\begin{equation}
\mathcal{A}_{n}=\epsilon \hbar n\text{ \ \ \ \ \ \ }(n=0,1,2.......),
\end{equation}

where $\mathcal{A}_{n}$\ is the area of the black hole horizon and $n$ is
the quantum number. $\epsilon $\ can be considered as a numerical
coefficient of the order of unity when all the fundamental constants except
Planck's constant ($\hbar $) have been set to one. In this regard, Hod \cite%
{Hod} made a semiclassical study, based on Bohr's correspondence principle
(the reader may refer to \cite{Bohr}), that the coefficient $\epsilon $\ can
be determined by manipulating the QNM frequencies of a vibrating black hole.
The Hod's result of $\epsilon $\ was also obtained by Kunstatter \cite%
{Kunstatter} who used the method of adiabatic invariant together with the
Bohr-Sommerfeld quantization condition. Essentially, Kunstatter showed that
a natural adiabatic invariant for system with energy $E$ and vibrational
frequency $\Delta \omega (E)$ is given by

\begin{equation}
I_{adb}=\int \frac{dE}{\Delta \omega (E)},\text{ \ \ }\Delta \omega =\omega
_{n+1}-\omega _{n}.
\end{equation}

and it has an equidistant spectrum $I_{adb}\simeq n\hbar $ at large quantum
numbers. Maggiore \cite{Maggiore} has recently extended the Kunstatter's
approach to determine the area spectrum of a black hole. According to the
Maggiore's argument, a black hole can be considered as a damped harmonic
oscillator. The proper frequency of the equivalent harmonic oscillator
corresponds to the QNM frequency ($\omega $), which plays an important role
in finding the entropy spectrum. Maggiore stated that the QNM frequencies
should be of the form $\omega =\left( \omega _{R}^{2}+\omega _{I}^{2}\right)
^{\frac{1}{2}}$, where $\omega _{R}$\ and $\omega _{I}$\ are the real and
imaginary parts of the frequency of the QNM. In the large $n$ limit, $\omega
_{I}\gg \omega _{R}$. Consequently one has to use $\omega _{I}$\ rather than 
$\omega _{R}$\ in the adiabatic quantity. Meanwhile, for a black hole $E$ is
identified with the mass ($M$) of the black hole in expression (2).

Inspired by Maggiore's idea Vagenas and Medved \cite{Vagenas, Medved}
obtained the area spectrum of rotating black holes. However, their results
showed that the area spectrum of a rotating black hole is not equidistant.
Similar to the studies of Vagenas and Medved, today one can see numerous
studies using Maggiore's proposal in the literature (see for instance \cite%
{Samp1,Samp2,Samp3,Samp4}). More recently, Wei et al. \cite{Wei} have
conjectured that for static chargeless black holes, which belong to
Einstein's gravity theory the spacing of both entropy and area spectra is
equidistant and independent of the dimension of spacetime. In the present
paper, we will examine whether Wei et al.'s conjecture is valid for higher
dimensional LDBHs or not.

The organization of the paper is as follows. Sect. 2 is devoted to a short
review of the higher dimensional LDBHs and their scalar perturbation. In
Sect. 3, we apply the semiclassical method to higher dimensional LDBHs and
obtain the entropy/area spectrum. Finally we give the concluding remarks.

\section{Massless Klein Gordon Equation in Higher Dimensional LDBHs}

In order to find the quantum of entropy (or area) spectrum by using
Maggiore's proposal, first we shall make the calculation of the massless
scalar wave equation on $N$-dimensional LDBHs and obtain the Zerilli type
wave equation \cite{Chandra}.

Let us first recall the $N$-dimensional ($N\geq 4$) LDBHs. As it was shown
in \cite{MSH}, their metrics are given by

\begin{equation}
ds^{2}=-fdt^{2}+\frac{dr^{2}}{f}+R^{2}d\Omega _{N-2}^{2},
\end{equation}

with the metric functions

\begin{equation}
f=\Sigma r\left[ 1-\left( \frac{r_{+}}{r}\right) ^{\frac{N-2}{2}}\right] ,%
\text{ \ }R=A\sqrt{r},
\end{equation}

It is obvious that metric (3) represents a static, non-rotating BH with a
horizon at $r_{+}.$ The constants $\Sigma $ and $A$ in the metric functions
(4) take different values according to the concerned theory (EMD, EYMD or
EYMBID) \cite{MSH}. For $r_{+}\neq 0$, the horizon hides the naked
singularity at $r=0$. However, in the extreme case of $r_{+}=0,$ the central
null singularity at $r=0$ is marginally trapped in which it does not allow
outgoing signals to reach external observers. Namely, even in the extreme
case of $r_{+}=0,$ metric (3) maintains its BH property.

As mentioned before here we consider a massless scalar field satisfying the
wave equation $\square \Psi =0$ in the higher dimensional LDBH spacetime (3)
where the metric functions to be used are chosen as in (4). In general, the
Laplacian operator on a $N$-dimensional metric is given by

\begin{equation}
\square =\frac{1}{\sqrt{-g}}\partial _{i}(\sqrt{-g}\partial ^{i}),\text{ \ \
\ }i=1...N,
\end{equation}

We look for a solution to this wave equation of the form,

\begin{equation}
\Psi=\rho(r)r^{\frac{2-N}{4}}e^{i\omega t}Y_{l}(\Omega_{N-2}),\text{ \ \ }%
Re(\omega)>0,
\end{equation}

in which $Y_{l}\left( \Omega _{N-2}\right) $ is the eigenfunction of $N-2$
dimensional Laplace-Beltrami operator $\nabla _{N-2}^{2}$ with the
eigenvalue $-l(l+N-3)$ \cite{Du}. After substituting harmonic eigenmodes (6)
into the wave equation (5) and making a straightforward calculation, one
obtains the following Zerilli equation \cite{Chandra},

\begin{equation}
\left[ -\frac{d^{2}}{dr^{\ast2}}+V(r)\right] \rho(r)=\omega^{2}\rho(r),
\end{equation}

where the effective potential \ is given by

\begin{equation}
V(r)=f(r)\left[ \frac{l(l+N-3)}{A^{2}r}+\frac{(N-2)(N-6)f(r)}{16r^{2}}+\frac{%
(N-2)f^{\prime}(r)}{4r}\right] ,
\end{equation}

and the tortoise coordinate is defined as,

\begin{equation}
r^{\ast }=\dint \frac{dr}{f(r)},
\end{equation}

which yields

\begin{equation}
r^{\ast }=\frac{2}{\Sigma (N-2)}\ln \left[ \left( \frac{r}{r_{+}}\right) ^{%
\frac{N-2}{2}}-1\right] .
\end{equation}

In principle Zerilli equation (7) can be solved with a particular set of
boundary conditions. It can be easily checked that the effective potential
(8) vanishes at the horizon ($r^{\ast }\rightarrow -\infty $) and becomes $%
\Sigma \left[ \frac{l(l+N-3)}{A^{2}}+\frac{\Sigma (N-2)^{2}}{16}\right] $\
at spatial infinity ($r^{\ast }\rightarrow \infty $). The latter results
will be used in the next section in order to obtain the QNMs of the LDBHs.

\section{Area/Entropy Quantization of Higher Dimensional LDBHs via QNM Method%
}

In this section, our main goal is to derive the area and entropy spectra of
the higher dimensional LDBHs by QNM method, which is prescribed by Maggiore 
\cite{Maggiore}. As described in \cite{Appro1,Appro2,Appro3}, here we use an
approximation method in order to define the QNMs. Since the effective
potential (8) vanishes at the horizon ($r^{\ast }\rightarrow -\infty $),
therefore the QNMs are defined to be those for which one has purely ingoing
plane wave at the horizon, namely,

\begin{equation}
\left. \rho(r)\right\vert _{QNM}\sim e^{i\omega r^{\ast}}\text{ at }r^{\ast
}\rightarrow-\infty,
\end{equation}

Now we will solve equation (7) in the near horizon limit and then impose the
above boundary condition to find the frequency of QNM i.e., $\omega $.

Expansion of the metric function $f(r)$ around the event horizon yields,

\begin{align}
f(r) & =f^{\prime}(r_{+}(r-r_{+})+O[(r-r_{+})^{2}]  \notag \\
& \simeq2\kappa(r-r_{+}),
\end{align}

where $\kappa $\ is the surface gravity, which is nothing but $\frac{1}{2}%
f^{\prime }(r_{+})$. Substituting this in the tortoise coordinate definition
(9) and evaluating the integration one finds,

\begin{equation}
r^{\ast}\simeq\frac{1}{2\kappa}\ln(r-r_{+}),
\end{equation}

Furthermore, substituting $\varepsilon =r-r_{+}$ in equations (8) and (12)
and performing Taylor expansion around $\varepsilon =0$\ we obtain the near
horizon form of the effective potential as,

\begin{equation}
V(\varepsilon)\simeq2\kappa\varepsilon\left[ \frac{l(l+N-3)}{A^{2}r_{+}}(1-%
\frac{\varepsilon}{r_{+}})+\frac{\left( N-2\right) \left( N-6\right) }{%
8r_{+}^{2}}\kappa\varepsilon+\frac{(N-2)\kappa}{2r_{+}}(1-\frac{\varepsilon 
}{r_{+}})\right] ,
\end{equation}

Also by substituting equation (13) into the Zerilli equation (7), we obtain
the near horizon form of the Zerilli equation:

\begin{equation}
-4\kappa^{2}\varepsilon^{2}\frac{d^{2}\rho(\varepsilon)}{d\varepsilon^{2}}%
-4\kappa^{2}\varepsilon\frac{d\rho(\varepsilon)}{d\varepsilon}+V(\varepsilon
)\rho(\varepsilon)=\omega^{2}\rho(\varepsilon),
\end{equation}

Solution of the above equation yields,

\begin{equation}
\rho(\varepsilon)\sim\varepsilon^{\frac{i\omega}{2\kappa}}U(\hat{a},\hat {b},%
\hat{c}),
\end{equation}

where $U(\hat{a},\hat{b},\hat{c})$\ is the confluent hypergeometric function 
\cite{Abramowitz}. The parameters of the confluent hypergeometric functions
are

\begin{align}
\hat{a} & =\frac{1}{2}+i(\frac{\omega}{2\kappa}-\frac{\hat{\alpha}}{\hat{%
\beta}\sqrt{\kappa}A}),  \notag \\
\hat{b} & =1+i\frac{\omega}{\kappa},  \notag \\
\hat{c} & =i\frac{\hat{\beta}\varepsilon}{2Ar_{+}\sqrt{\kappa}},
\end{align}

where

\begin{align}
\hat{\beta}& =\sqrt{8l(l+N-3)-\kappa A^{2}(N-10)(N-2)},  \notag \\
\hat{\alpha}& =l(l+N-3)+\frac{N-2}{2}\kappa A^{2},
\end{align}

In the limit of $\varepsilon \ll 1$, the solution (16) reduces to the form

\begin{equation}
\rho(\varepsilon)\sim C_{1}\varepsilon^{-\frac{i\omega}{2\kappa}}\frac {%
\Gamma(i\frac{\omega}{\kappa})}{\Gamma(\hat{a})}+C_{2}\varepsilon ^{\frac{%
i\omega}{2\kappa}}\frac{\Gamma(-i\frac{\omega}{\kappa})}{\Gamma (1+\hat{a}-%
\hat{b})},
\end{equation}

where constants $C_{1}$ and $C_{2}$ denote the amplitudes of the
near-horizon outgoing and ingoing waves, respectively. Now, since there is
no outgoing wave in the QNM (11) at horizon, the first term should be
vanished. This will happen at poles of the Gamma function of the denominator
of the first term. The poles of the Gamma function will definitely determine
the frequencies of the QNMs. Thus, one can read the frequencies of the QNMs
of the higher dimensional LDBHs as,

\begin{equation}
\omega_{n}=\frac{2\sqrt{\kappa}\hat{\alpha}}{\hat{\beta}A}+i(2n+1)\kappa,
\end{equation}

where $n=1,2,3....$. It is very important to note that the existence of the
frequencies $\omega_{n}$\ strictly depends on $\hat{\beta}$ parameter$.$ It
must certainly be real, which brings us a condition as follows

\begin{equation}
8l(l+N-3)>\kappa A^{2}(N-10)(N-2),\text{ \ \ (for }N\geq 11\text{)},
\end{equation}

Whenever the above condition holds, the imaginary part of the frequency of
the QNM is

\begin{equation}
\omega _{I}=(2n+1)\kappa =\frac{2\pi }{\hbar }(2n+1)T_{H},  \label{Star}
\end{equation}

where $T_{H}=\frac{\hbar \kappa }{2\pi }$ is so-called the Hawking
temperature.

Now the energy of the such NAF black holes is calculated by the quasilocal
mass definition \cite{BrownYork}. Thus, one can determine the quasilocal
mass of the higher dimensional LDBHs as,

\begin{equation}
M=\frac{r_{+}^{\frac{N-2}{2}}\left( N-2\right) \Sigma A^{N-2}}{8},
\end{equation}

Since from (22) $\Delta \omega =\omega _{n+1}-\omega _{n}=4\pi T_{H}$, the
adiabatic invariant quantity (2) in this case yields,

\begin{equation}
I_{adb}=\frac{\hbar }{4\pi }\int \frac{dM}{T_{H}},
\end{equation}

By considering the first law of thermodynamics, $T_{H}dS_{BH}=dM$, one can
easily see that

\begin{equation}
I_{adb}=\frac{S_{BH}}{4\pi }\hbar ,
\end{equation}

Finally, according to the Bohr-Sommerfeld quantization rule

\begin{equation}
I_{adb}=\hbar n,
\end{equation}

one gets the spacing of the entropy spectrum as

\begin{equation}
S_{n}=4\pi n,
\end{equation}

Recalling the relation $S=\frac{\mathcal{A}}{4},$ the area spectrum is given
by

\begin{equation}
\mathcal{A}_{n}=16\pi n,
\end{equation}

with the spacing

\begin{eqnarray}
\Delta S &=&S_{n+1}-S_{n}=4\pi  \notag \\
&=&\frac{\Delta \mathcal{A}}{4}.
\end{eqnarray}

We remark that both the entropy and area spectra are equally spaced and
independent of the dimension of spacetime. This result is in agreement with
that of Wei et al.'s conjecture \cite{Wei}. As mentioned before, the above
result is valid only for large quantum numbers $n$ and the fulfilling of the
condition (21).

\section{Conclusion}

In this paper we investigated the area and entropy spectra of higher
dimensional LDBHs. By utilizing the QNM method, it is showed the QNM
frequencies become apparent with a particular condition $8l(l+N-3)>\kappa
A^{2}(N-10)(N-2)$ when $N$ becomes greater than and equal to 11D. Unless
this condition fails, the higher dimensional LDBHs have equally spaced area
and entropy spectra. Both spectra are independent of the dimension of
spacetime which means that our results in accordance with the conjecture of 
\cite{Wei}. As a final remark, one should keep in mind that all calculations
made here are semiclassical and based on Bohr-Sommerfeld quantization
condition and QNMs. Finally, we want to point out that since the LDBHs are
conformally related to the Brans-Dicke BHs \cite{BDBhs}, the same analysis
might work for those BHs as well. This is going to be our next problem in
the near future.

\end{document}